# Single Phase Synthesis and Room temperature Neutron Diffraction Studies on Multiferroic $PbFe_{0.5}Nb_{0.5}O_3$

*Shidaling Matteppanavar [1], Basavaraj Angadi [1,\*], Sudhindra Rayprol[2]*

[1]Department of Physics, JB Campus, Bangalore University, Bangalore –560056
[2]UGC-DAE-CSR, Mumbai Centre, BARC, Mumbai – 400085
\*Corresponding Author, e-mail : brangadi@gmail.com

**Abstract.** The Lead-iron-niobate, ($PbFe_{0.5}Nb_{0.5}O_3$ or PFN) was synthesized by low temperature sintering Single Step / Solid State Reaction Method. The 700 °C / 2 hrs calcined powder was sintered at 1050 °C / 1 hr. The sintered pellets were characterized through X-Ray Diffraction and Neutron Diffraction at room temperature. It is found from the XRD pattern that the materials is in single phase with no traces of pyrochlore phase. It was also confirmed from the neutron diffraction pattern, the structure of PFN to be monoclinic, belongs to space group *C m*. Structural studies has been carried out by refining the obtained neutron diffraction data by Reitveld refinement method using *Full Prof Program*. The neutron diffraction pattern at 290K (near room temperature) was selected to refine the structure. The Lattice parameters obtained are; a = 5.6709 Å, b = 5.6732 Å, c = 4.0136 Å, and α = 90, β = 89.881, γ = 90.000. The P-E measurements showed hysteretic behavior with high remnant polarization.

**Keywords:** Multiferroics, $PbFe_{0.5}Nb_{0.5}O_3$ (PFN), Neutron Diffraction, Ferroelectric hysteresis.
**PACS:** 75.85.+t ; 61.05.C- ; 61.05.F- ; 77.80.Dj

## INTRODUCTION

Multiferroics are relatively new class of materials with ferroelectric and ferromagnetic (anti-ferromagnetic) orderings coexisting in a certain temperature [1-2]. $PbFe_{0.5}Nb_{0.5}O_3$ (PFN) multiferroics has the complex perovskite structure and was discovered and synthesized at the end of 1950s [5]. Its ferroelectric Curie temperature ($T_C$) is around 383 K and the anti-ferromagnetic order begins at a Néel temperature of $T_N$=143 K [1, 3－6]. It is well reported in the literature that synthesis of single phase PFN with perovskite phase has been difficult by the conventional methods, due to the formation of unwanted pyrochlore phases. Other alternative methods either involve two or more calcination steps or high sintering temperatures. In this work, we employed a single-step calcination and a low temperature sintering method to achieve single phase PFN.

## EXPERIMENTAL

The single phase PFN was synthesized through the single step reaction method [4]. The reagent grade $Pb(NO_3)_2$, $Fe_2O_3$ and $Nb_2O_5$ were taken stoichiometrically and ground in agate pestle and mortar in ethanol medium for 2 hours. The dried powder was calcined at 700 °C 2 hours. After the calcination stage the powder was ground again and polyvinyl alcohol (PVA) was added. Pellets of 10mm (5 mm) in diameter and 2-3mm thickness were uniaxially pressed at 50 kN using a hydraulic press. The pellets were sintered at 1050 °C for 1 hour in a closed Pb rich environment environment to minimize the PbO evaporation. The PFN samples were characterized by X-ray powder diffraction using Phillips (1070 model) diffractometer with Cu $K_α$ (1.5406 Å) radiation.

Neutron diffraction measurements were carried out on the PSD-based powder diffractometer at Dhruva reactor, BARC, using a wavelength of 1.48 °A. For structural analysis Rietveld refinement was carried out through the *FullProf* program. P- E measurements were done using sawer-tower circuit. The bulk density was measured using Archimedes principle with distilled water.

## RESULTS AND DISCUSSION

Figure 1 shows the XRD pattern of sintered PFN sample showing single phase and matching well with the JCPDS pattern no. 32-0522. The formation of unwanted pyrochlore phases was completely bypassed in this Single step method and with the 100% perovskite phase. The low temperature calcination along with the closed and Pb rich environment helps in minimizing the PbO evaporation, which has been known to be the origin for the pyrochlore phase formation. Various calcination and sintering temperatures and time duration were tried to achieve single phase PFN, but it was observed that 700 °C/2 hrs calcination and 1050 °C/1 hr sintering were optimum conditions.

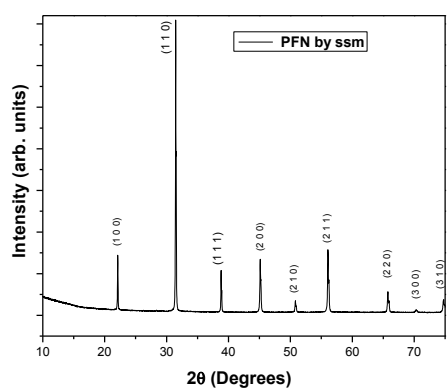

**FIGURE 1.** X-ray diffraction pattern of PFN sample.

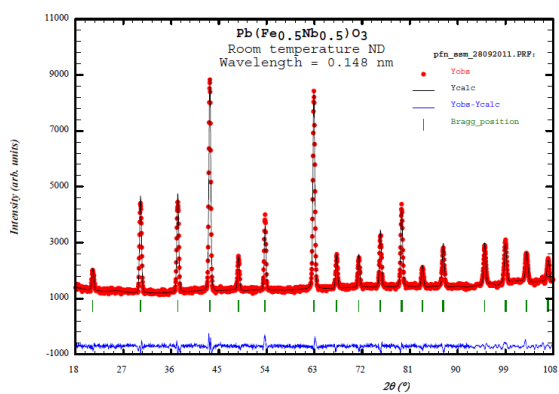

**FIGURE 2.** Neutron Diffraction data of PFN at 300K

The room temperature Neutron diffraction data of the sintered PFN is shown in Fig.2. The data was refined with monoclinic structure with space group *Cm*. Obtained R-factors for the monoclinic phase are $R_p$=2.87 ; $R_{wp}$= 3.71; $R_{exp}$= 2.47; Chi2: 2.26. The Lattice parameters obtained are; a = 5.6709Å, b = 5.6732 Å, c = 4.0136 Å, and α = 90, β = 89.881, γ = 90. The density of the sintered pellets measured using Archimedes method with distilled water as displacement liquid is around 8.465 g/cm$^3$. This shows the dense compaction of the sintered pellets.

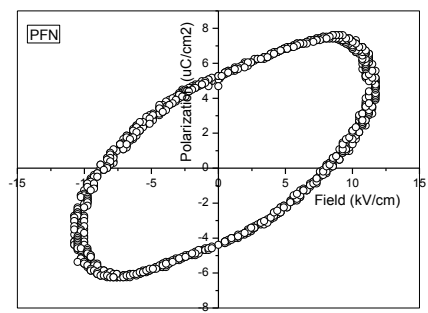

**FIGURE 3.** P-E Hysteresis loop for the sintered PFN

The P-E hysteresis loop for the PFN is shown in Fig. 3, which clearly indicates the typical ferroelectric nature. The observed remnant polarization and coercive field values are $P_R$≈ 5 μC/cm$^2$ and $E_C$ ≈ 8 kV/cm, respectively.

In conclusion, the multiferroic PFN was synthesized by Single step method using lower calcination and sintering temperatures. The 100% perovskite phase with no traces of pyrochlore phase was obtained. The room temperature XRD and Neutron diffraction studies showed that the synthesized PFN is in the monoclinic phase with space group *Cm*. The samples exhibit typical ferroelectric P-E hysteresis with the obtained values of $P_R$≈ 5 μC/cm$^2$ and $E_C$ ≈ 8 kV/cm.

## ACKNOWLEDGMENTS

The authors thank the UGC-DAE-CSR, Mumbai Centre for the financial support through CRS-M 159.